\documentclass{pkas}

\def\year{---} % publication year
\def\month{---} % publication month
\def\volume{---} % journal volume
\def\issue{---} % journal issue
\def\beginpage{0} % first page of article
\def\endpage{99} % last page of article
\def\received{---} % date paper was received by PKAS
\def\accepted{---} % date of acceptance
\date{Received \received ; accepted \accepted}

\title{
Initial analysis of extragalactic fields using a new AKARI/IRC analysis pipeline
}

\author[1]{Helen~Davidge}
\author[1]{Stephen~Serjeant}
\author[1,2,3]{Chris~Pearson}
\affil[1]{Department of Physical Sciences, The Open University, Milton Keynes, MK7 6AA, UK; \email{helen.davidge@open.ac.uk}}
\affil[2]{RAL Space, Rutherford Appleton Laboratory, Chilton, Didcot, Oxfordshire, OX11 0QX, UK; \email{chris.pearson@stfc.ac.uk}}
\affil[3]{Oxford Astrophysics, University of Oxford, Oxford, OX1 3RH, UK
 \email{ }}
\def\corrauthor{
H. Davidge
}

\def\runningauthor{
Helen Davidge et al.
}

\def\runningtitle{
Initial analysis of extragalactic fields using a new AKARI/IRC analysis pipeline
}

\def\keywords{
Methods: data analysis, Infrared: source counts
}

\def\abstracttext{
We present the first results of a new data analysis pipeline for processing extragalactic AKARI/IRC images. The main improvements of the pipeline over the standard analysis are the removal of Earth shine and image distortion correction. We present the differential number counts of the AKARI/IRC S11 filter IRAC validation field. The differential number counts are consistent with S11 AKARI NEP deep and 12$\mu$m WISE NEP number counts, and with a phenomenological backward evolution galaxy model, at brighter fluxes densities. There is a detection of deeper galaxies in the IRAC validation field.
}

\begin{document}
\pkashead %% set title, authors, abstract, etc.

%%%%%%%%%%%%%%%%%%%%%%%%%%%%%%%%%%%%%%%%%%%%%%%%%%%%%%%%%%%%%%%%%%%%%
%%% BEGIN MAIN TEXT HERE %%%%%%%%%%%%%%%%%%%%%%%%%%%%%%%%%%%%%%%%%%%%
%%%%%%%%%%%%%%%%%%%%%%%%%%%%%%%%%%%%%%%%%%%%%%%%%%%%%%%%%%%%%%%%%%%%%

\section{Introduction}

There is currently no satellite observing in the mid-infrared, and none will be operational until 2018 at the earliest; the projected launch date of James Webb Space Telescope (JWST). Therefore best use must be made of existing archive data. The Infrared Camera (IRC) \citep{b2}, one of the two instruments on board the AKARI satellite \citep{b3} has a large amount of archive data. The IRC consisted of 9 photometric bands from 2-24$\mu$m. AKARI/IRC covers the wavelength desert between the Spitzer IRAC and MIPS instruments from 8-24$\mu$m \citep{b4}. This paper reports on a new optimised general data reduction pipeline specifically tailored for extragalactic fields. Section 2 gives an overview of the improved pipeline. Section 3 shows initial results of analysis of one the AKARI/IRC deepest extragalactic fields, the IRAC validation field. This field was observed multiply by Spitzer space telescope and the Herschel space observatory. The field was also used by both telescopes as a calibration field. Section 4 discuses the preliminary results, and Section 5 states the conclusions.

\section{IRC data reduction}
\subsection{Overview}

\begin{figure}
\centering
\includegraphics{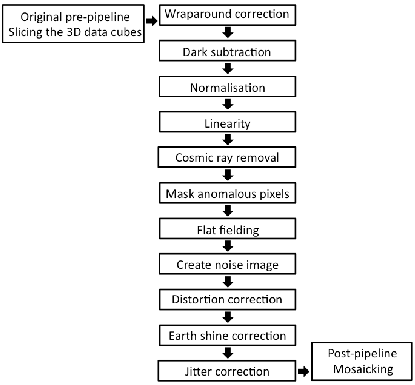}
\caption{Outline of new optimised pipeline steps}
\label{fig:Outline_of_IDL_pipeline_steps}
\end{figure}

There is a standard pipeline toolkit to process and analyse IRC images written by the AKARI instrument team \citep{b4}. We have developed a pipeline optimised for processing extragalactic images to extract the highest signal to noise point sources in deep field images. Figure \ref{fig:Outline_of_IDL_pipeline_steps} summaries the steps of the new pipeline. The new pipeline includes modules for improving cosmic ray detection, Earth shine removal, image distortion correction and astrometry correction. 

\subsection{Removal of Earth shine}

 A major problem with many of the IRC observations is that the images are severely affected by reflected Earth shine. This creates large-scale features across the image. The gradient of the large-scale features varies during a pointing. The Earth shine is removed from each image by fitting a boxcar median to the background flux of each image, and subtracting this from the original image.

\subsection{Image distortion correction}

The raw IRC images are geometrically optically distorted. The new pipeline has a unique distortion polynomial for each of the nine filters. The new pipeline maps the detector onto the sky using the correct distortion polynomial to create a dewarped image. 

\section{Results}

\begin{figure}
 \centering
 \includegraphics{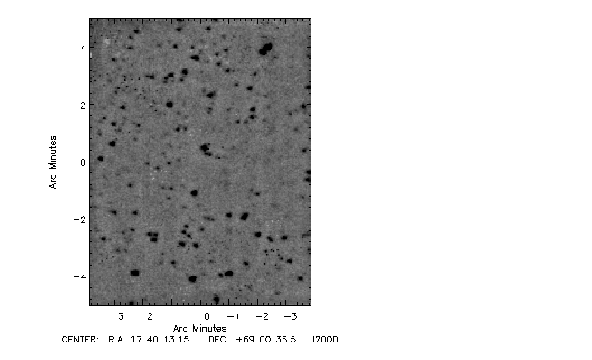}
 \caption{Image of the IRAC validation field - S11 filter}
 \label{fig:coadd}
\end{figure}

\begin{figure}
 \centering
 \includegraphics{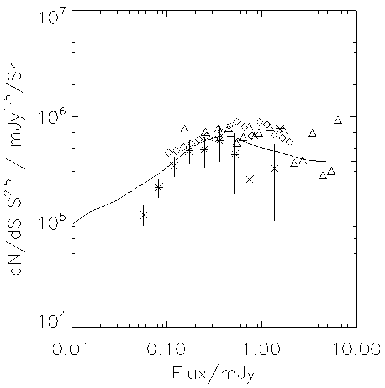}
 \caption{Differential number counts. Stars are the AKARI S11 IRAC validation field raw number counts. Diamonds are AKARI IRC S11 NEP deep, triangles are 12$\mu$m WISE NEP, the line is a phenomenological backward evolution galaxy model of \citet{b7}}
 \label{fig:number_counts}
\end{figure}

The first extragalactic  field to be processed by the new pipeline is the IRAC validation field (IRAC VF). The mosaicked S11 filter image is shown in figure \ref{fig:coadd}. We present the galaxy number counts of the S11 filter data. Source extraction was carried out using a optimal point source filer technique \citep{b10} that maximises the signal to noise for isolated point sources on a flat background. For the error in number counts for bins which contain small numbers of galaxies, the method from \citet{b9} was used. Fluxes were calibrated using the IRC manual flux calibration \citep{b4}. Figure \ref{fig:number_counts} shows the raw differential galaxy number counts plotted with the S11 AKARI NEP (north ecliptic pole) deep number counts \citep{b11}, the 12$\mu$m WISE NEP number counts \citep{b12} and the phenomenological backward evolution model of \citet{b7}. The model is multi-component and defined in the far-infrared, consisting of normal, star-forming and AGN populations. The number counts have not been corrected for completeness and reliability. These corrections should be small for the brighter 5$\sigma$ selected sample but may be more significant at the faint end of the source counts. For a more detailed analysis of the number counts, and analysis of other IRC filters see \citet{b8}.

\section{Discussion}

The raw number counts for the IRAC VF from 0.1 to 0.4 mJy fit the model well. The IRAC VF has a smaller survey area than the NEP and thus the number counts have larger errors. Figure \ref{fig:number_counts} shows the IRAC VF number counts goes deeper than the AKARI NEP deep number counts. Correct removal of the large scale structure should improve the number counts, especially for fainter fluxes and could yield even deeper point sources.

\section{Conclusions}

The number counts for the S11 IRAC VF, S11 AKARI NEP, 12$\mu$m WISE NEP and model agree fairly well at higher fluxes. The IRAC VF number counts are smaller at fainter fluxes, probably due to incompleteness in the galaxy distribution. Using the new pipeline to process the S11 IRAC VF, we have found fainter point sources and hence gone deeper than the AKARI NEP survey.

%%% ACKNOWLEDGMENTS (IF ANY) %%%%%%%%%%%%%%%%%%%%%%%%%%%%%%%%%%%%%%%%

\acknowledgments

The AKARI Project is an infrared mission of the Japan Space Exploration Agency (JAXA) Institute of Space and Astro-nautical Science (ISAS), and is carried out with the participation of mainly the following institutes; Nagoya University, The University of Tokyo, National Astronomical Observatory Japan, The European Space Agency (ESA), Imperial College London, University of Sussex, The Open University (UK), University of Groningen / SRON (The Netherlands), Seoul National University (Korea). The far-infrared detectors were developed under collaboration with The National Institute of Information and Communications Technology.

%%% APPENDICES (IF ANY) %%%%%%%%%%%%%%%%%%%%%%%%%%%%%%%%%%%%%%%%%%%%%

%%% CALL LIST OF REFERENCES (natbib STYLE) %%%%%%%%%%%%%%%%%%%%%%%%%%

\end{document}